\documentclass[12pt,usenatbib]{mn2e}
\usepackage{graphicx}
\usepackage{epsf}
\def\lsim{\mathrel{\lower0.6ex\hbox{$\buildrel {\textstyle <}
 \over {\scriptstyle \sim}$}}}
\def\gsim{\mathrel{\lower0.6ex\hbox{$\buildrel {\textstyle >}
 \over {\scriptstyle \sim}$}}}

\def\rvir{\mathrel{\textit{r}}_{\rm vir}}

\def\asat{\mathrel{\hat{\rm a}_{\rm sat}}}
\def\bsat{\mathrel{\hat{\rm b}_{\rm sat}}}
\def\csat{\mathrel{\hat{\rm c}_{\rm sat}}}

\def\lsim{\mathrel{\lower0.6ex\hbox{$\buildrel {\textstyle <}
 \over {\scriptstyle \sim}$}}}
\def\gsim{\mathrel{\lower0.6ex\hbox{$\buildrel {\textstyle >}
 \over {\scriptstyle \sim}$}}}

\begin{document}

\title[Alignment of Milky Way satellites]{How common is the Milky Way -
satellite system alignment?} 
\author[N. I. Libeskind et al.]{
\parbox[h]{\textwidth}{Noam I Libeskind$^{1,2}$, Carlos S Frenk$^3$, Shaun
Cole$^3$, Adrian Jenkins$^3$ \& John C Helly$^{3}$.}
\vspace{6pt} \\
$^{1}$Racah Institute of Physics, Hebrew University of Jerusalem, Givat Ram,
Jerusalem, Israel, 91904\\ 
$^{2}$Astrophysikalisches Institut Potsdam, An der Sternwarte 16, Potsdam 14482, Germany\\
$^{3}$Department of Physics, University of Durham, Science
  Laboratories, South Road, Durham, DH1 3LE, U.K.\\}

%\date{Accepted 1988 December 15. Received 1988 December 14; in original form

 %1988 October 11} 

%\pagerange{\pageref{firstpage}--\pageref{lastpage}} \pubyear{2002}

\maketitle \begin{abstract} The highly flattened distribution of satellite galaxies in the Milky Way presents a number of puzzles. Firstly, its polar alignment stands out from the planar alignments commonly found in other galaxies. Secondly, recent proper motion measurements reveal that the orbital angular momentum of at least 3, and possibly as many as 8, of the Milky Way's satellites point (within 30 degrees) along the axis of their flattened configuration, suggesting some form of coherent motion. In this paper we use a high resolution cosmological simulation to investigate whether this pattern conflicts with the expectations of the cold dark matter model of structure formation. We find that this seemingly unlikely set up occurs often: approximately $35$\% of the time we find systems in which the angular momentum of 3 individual satellites point along, or close to, the short axis of the satellite distribution. In addition, in $30$\% of the systems we find that the net angular momentum of the 6 best aligned satellites lies within 35 degrees of the short axis of the satellite distribution, as observed for the Milky Way.

\end{abstract}

%\keywords{cosmology: theoryÑdark matterÑgalaxies: haloesÑ galaxies: formationÑ
%gravitation}

\section{Introduction}

The system of bright satellites orbiting in the halo of the Milky Way
has a number of properties that seem unusual. Most of these satellites
define a tight plane in the sky whose normal is roughly perpendicular
to the Galaxy's spin angular momentum vector
\citep[]{Lyndenbell76,Kunkel76,Lyndenbell82}. In addition, this
so-called \textit{disc-of-satellites} \citep[]{Kroupa05} appears to be well aligned with
the orbital angular momentum of individual satellites, indicating some form of
coherent motion \citep{Metz08}.

The observed, highly anisotropic set-up is not an obvious outcome of
the $\Lambda$CDM model in which the Universe is assumed to be composed
primarily of cold dark matter (CDM) and vacuum energy associated with
a cosmological constant ($\Lambda$). In this paradigm, structures form
hierarchically via the merging of small subgalactic units formed out
of the unstable overdense peaks of the initial matter density
field. Satellite galaxies are identified with the substructures that
survive the violent gravitational processes associated with
hierarchical growth. High resolution N-body simulations of galaxy
formation have shown that small bound subhalos can survive these
merging processes and that many remain as distinct substructures
embedded in a present day galactic dark matter halo
\citep[e.g.][]{Klypin99, Moore99,Diemand07, Springel08}.

Before the flood of discoveries of new Milky Way satellites in the Sloan Digital
Sky Survey (SDSS) \citep[]{Willman05,Belokurov06a, Belokurov06b, Zucker06a,
Zucker06b}, it had been known for at least 3 decades that rather than being
isotropically distributed in space, the 11 most luminous `classical' satellite
galaxies of the Milky Way are all aligned along a great circle on the sky
\citep[]{Lyndenbell76,Kunkel76}. \cite{Metz09} have argued that the
newly discovered SDSS satellites also lie on or close to this great
circle. This `galactic pancake' \citep[]{Libeskind05} of satellites
was found to be aligned roughly perpendicular to the plane of the disc
of the Milky Way \citep{Lyndenbell82}. This particular set-up may be
related to the ``Holmberg effect,'' the tendency for satellites of
extragalactic systems to be preferentially distributed along a
specific direction. A number of studies
\citep{Holmberg69,Zaritsky97,SalesLambas04,Brainerd05,Yang05} have
detected some alignment of satellite galaxy systems relative to their
hosts, but disagree on the nature of its
orientation. \cite{Holmberg69} and
\cite{Zaritsky97} found  ``Milky Way'' type alignments (i.e. polar, with satellites
avoiding the equatorial regions), while planar alignments (i.e. with satellites
avoiding the polar regions) have been observed in the 2-degree-field Galaxy
Redshift Survey \citep[2dFGRS,][]{Colless01} by
\cite{SalesLambas04} and, in the SDSS, by \cite{Brainerd05} and \cite{Yang05}.

\cite{Kroupa05} argued that the anisotropic nature of the distribution of the
Milky Way's satellite population contradicts the expectations of the
$\Lambda$CDM model since the simple hypothesis that satellite galaxies
are a random subsample of the entire subhalo population would imply
near isotropy. However, recent numerical studies of the formation of
satellites in the $\Lambda$CDM cosmology
\citep{Libeskind05,Libeskind07,Kang05,Zentner05} have found that there
is not a one-to-one match between the brightest satellites and the
most massive subhalos and indeed that the satellite galaxies are not a
random subsample of the background subhalo population. Instead, the
most luminous satellites populate the subset of subhalos which were
the most massive before they were accreted. These studies also
suggested that the tendency for satellites to align themselves in a
great pancake may be a generic result of infall along filaments that
form in the hierarchical buildup of a galactic halo or of accretion in
groups \citep{Li08}. This kind of coherent motion should perhaps also be reflected in the kinematics of
the satellite galaxies.

Proper motions for some of the Milky Way's satellite galaxies (at the level of
milliarcsec/yr) have now been measured using both ground-
\citep{Scholz94,Ibata97,Dinescu04,Dinescu05,Schweitzer95,Schweitzer97} and
space-based telescopes
\citep[i.e. HST][]{Kallivayalil06a,Kallivayalil06b,Piatek03,Piatek05,Piatek06,Piatek07}.
Using these data, \cite{Metz08} calculated the orbital angular
momentum of 6 satellites. Remarkably, they found the angular momentum
vectors of the satellites to be well aligned, generally pointing in
the same direction, within 30 degrees, on the sky. This effect was not
seen in the simulations of
\citep{Libeskind05,Libeskind07} which were shown to be consistent with a random
orientation for the satellites' angular momenta. This conclusion, however, was
based on a very small sample of simulations of Milky-Way systems.

In this study we use a high resolution simulation of a large cosmological volume
to explore the alignment of the positions and angular momenta of satellites in
dark matter halos. With a much large sample than those previously considered, we
are able to investigate these properties with a high degree of statistical
significance.

\section{Methods}
\subsection{{N}-body simulation and semi-analytic model}

The parameters of the hMS $\Lambda$CDM N-body simulation that we
analyze (listed in Table~1) are the same as the \textit{Millennium Simulation} \citep{Springel05a} except that the box size is smaller and the mass resolution approximately ten times better  
\citep[see][for details]{Gao08}. 
The simulation was performed using the
\textsc{L-Gadget2} code \citep{Springel05} on 128 processors at the
Cosmology Machine of Durham University's Institute for Computational
Cosmology. The simulation was populated with `galaxies' using the
semi-analytical model of \cite{Bower06} applied to halo merger trees
extracted from the simulation.

\begin{table}
\begin{center}
\caption{Parameters of the hMS N-body simulation. (1) Box size (in $h^{-1}$ Mpc);
(2) number of particles; (3) dark matter particle mass (in $
{\rm M}_{\odot}$); (4) mean matter density; (5) vacuum energy density; (6) Hubble constant
(in units of 100 ${\rm km\,s^{-1}Mpc^{-1}}$); and (7) normalization of the matter power
spectrum.} 
 \begin{tabular}{l l l l l l l}
         $l_{\rm box} $  & N  &  $ m_{\rm dm}$ & $\Omega_{\rm m}$
        & $\Omega_{\Lambda}$ & $h$ & $\sigma_{8}$\\
   \hline
   \hline
	100 & 7.29$\times10^{8}$ & 1.3$\times10^{8} $ &0.25 & 0.75& 0.73 & 0.9 \\
    \hline
    \hline
 \end{tabular}
 \end{center}
\label{sim_table}
\end{table}

Large high-resolution simulations such as ours resolve a wealth of
gravitationally bound substructures inside larger objects. When a small halo is
accreted by a larger one, tidal stripping causes the substructure to lose mass,
while dynamical friction against the parent halo's background material causes
its orbit to spiral in towards the centre. In order to identify and follow these
accreted substructures we use the algorithm \textsc{Subfind} \citep{Springel01},
which uses an excursion set approach recursively to identify local maxima in the
dark matter density field. The extent of a substructure is then determined by
locating saddle points in the dark matter density field. The gravitational
binding energy is calculated and unbound particles are rejected. Finally a 20
particle lower limit is imposed on the substructures.

We identify subhalos at all redshifts in the simulation and use them
to build halo 
merger trees which describe the hierarchical build up of structures 
\citep[see][for a full description]{Harker06}. The 
trees provide the backbone for modelling the evolution of the
baryons. We use the Durham semi-analytic galaxy formation model,
\textsc{Galform} \citep{Cole00,Benson02, Baugh05}, as extended by \cite{Bower06}
which, in brief, includes the following physical processes: (i)
shock-heating and virialization of halo gas; (ii) radiative cooling of
hot halo gas onto a galactic disc; (iii) star formation from cool gas
in the disc and during starbursts in the bulge; (iv) the evolution of
stellar populations; (v) photoionization and its effect on the thermal
properties of the intergalactic medium; (vi) reheating and expulsion
of cooled gas due to feedback processes associated with supernovae,
stellar winds and AGN; (vii) chemical evolution of gas and stars;
(viii) reddening due to dust absorption; (ix) galaxy mergers; (x)
formation of black holes in the bulges of galaxies via the accretion
of gas during mergers and disc instabilities; (xi) the evolution of
the size of bulges and discs;

\subsection{The satellite sample}
\label{sample.def}

The semi-analytic model produces a galaxy catalogue at each simulation
output. Resolution tests, in which we increased the minimum halo mass
used in the merger trees, indicate that our catalogues are complete
for satellite galaxies with \textit{V}-band magnitude brighter than
-8. In the Milky Way, this magnitude limit corresponds to the 11th
brightest satellite, \citep[Draco, which has $M_{\rm V} = -8.8$;
see][]{Mateo98}. Although the number of known dwarf galaxies orbiting
in the Milky way halo has increased by at least 15 in the past 3
years, as new local group dwarf satellites have been discovered in the
SDSS, all of the newly discovered galaxies are faint and fall below
our magnitude limit. For this reason, we do not consider them in this
paper. (Note that the newly discovered SDSS satellites are spatially
biased because of the limited sky coverage of the SDSS so it is
unclear how their properties compare with those of the classical
satellites.) Thus, when comparing with observations of the Milky Way
satellite system we only consider the 11 classical satellite galaxies
whose luminosity is greater than our magnitude limit.

In order to identify central galaxies that resemble the Milky Way, we apply a
halo mass cut which excludes galaxies residing in haloes of mass $\leq 2\times
10^{11}$~M$_{\odot}$ and $\geq 2\times 10^{12}$~M$_{\odot}$. To refine further the
search for Milky Way-type galaxies, we also apply a luminosity cut to eliminate
central galaxies fainter than $-20$ in the $V$ band. We consider only systems
that contain 11 or more satellites within the halo virial radius, $\rvir$, defined as 
the distance from halo centre within which the mean interior density is 200 times the
critical density.

\begin{figure}
\includegraphics[width=20pc]{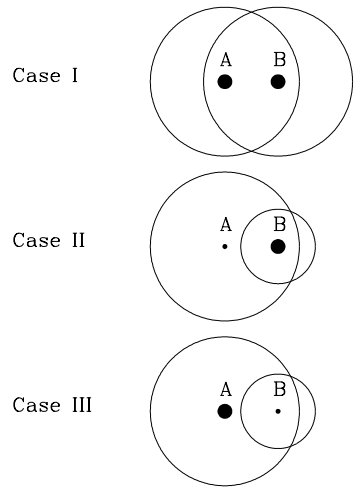}
\caption{Isolation criteria. In case I (\textit{top}), galaxy A is within galaxy
B's virial radius and galaxy B is within galaxy A's virial radius; in this case
both galaxies are eliminated from the sample. In case II (\textit{middle}),
galaxy B lies within galaxy A's virial radius (but not vice-versa) and galaxy B
is more massive than galaxy A; in this case both galaxies are eliminated from
the catalogue. In case III (\textit{bottom}), galaxy~B is within galaxy A's
virial radius and is less massive than galaxy A; in this case galaxy~B is
regarded as a satellite of galaxy A.}
\label{isolate}
\end{figure}

In order to prevent contamination of an alignment signal by nearby
central galaxies, we
impose the isolation criteria shown schematically in Fig.~\ref{isolate}. We
eliminate for the galaxy catalogue any two central galaxies A and B that lie within each
other's virial radius (Case I). If galaxy B lies within the virial radius of
another galaxy A, we eliminate both from our sample if B has a stellar
mass greater than that of A
(Case II), and turn B into a satellite of A if it is less massive than A (Case
III). In practice, cases I, II and III are rare with only 29, 63 and 102 galaxies
falling into these categories respectively.

When applied to the galaxy list produced by the semi-analytic model, the halo
mass cut leaves us with 30\,946 systems. The central galaxy magnitude
cut reduces our sample to 3\,230 halos, while the isolation criteria
leaves us 3\,201 systems. This sample is further reduced to 436
galaxies which posses at least 11 satellites. The median halo masses of
each sample after these 4 cuts are $10^{11.32}$,
$10^{11.82}$, $10^{11.82} $ and $10^{11.96} {\rm M}_{\odot}$ respectively.
Thus the median halo mass is 
only increased by 38\% by the requirement that each system host at
least 11 satellites.

\section{Satellite galaxy alignments}

We first consider the alignment between satellite galaxy systems and
the shape of the parent halo and then investigate the alignment of the
satellites' orbital angular momentum vectors.

\subsection{Orientation of the halo and satellite positions}
\label{halo.oreintation}
Once the isolation criteria, and mass and luminosity cuts have been applied we
obtain a list of central galaxies. We wish to examine the flattening
of both their haloes and  satellite systems. To do so, we first define the
inertia, or second moments tensor,
\begin{equation}
I_{j,k}=\sum_{i} x_{j,i}x_{k,i} ,
\label{inertia}
\end{equation}
where we sum either over all the particles or all the satellites in
the halo. We diagonalize this tensor and obtain the principal axes of
the system, defined conventionally such that $a > b > c$. Previous
studies \cite[e.g.][]{Bullock00,Libeskind05,Libeskind07} have shown
that although haloes tend to be mildly aspherical, their satellite
populations exhibit a much more pronounced flattening. In order to
verify this with our sample, we calculate the eigenvalues of the
inertia tensor for the dark matter particles in each halo and for the
systems of 11 brightest galaxies within $\rvir$.

\begin{figure}
\includegraphics[width=20pc]{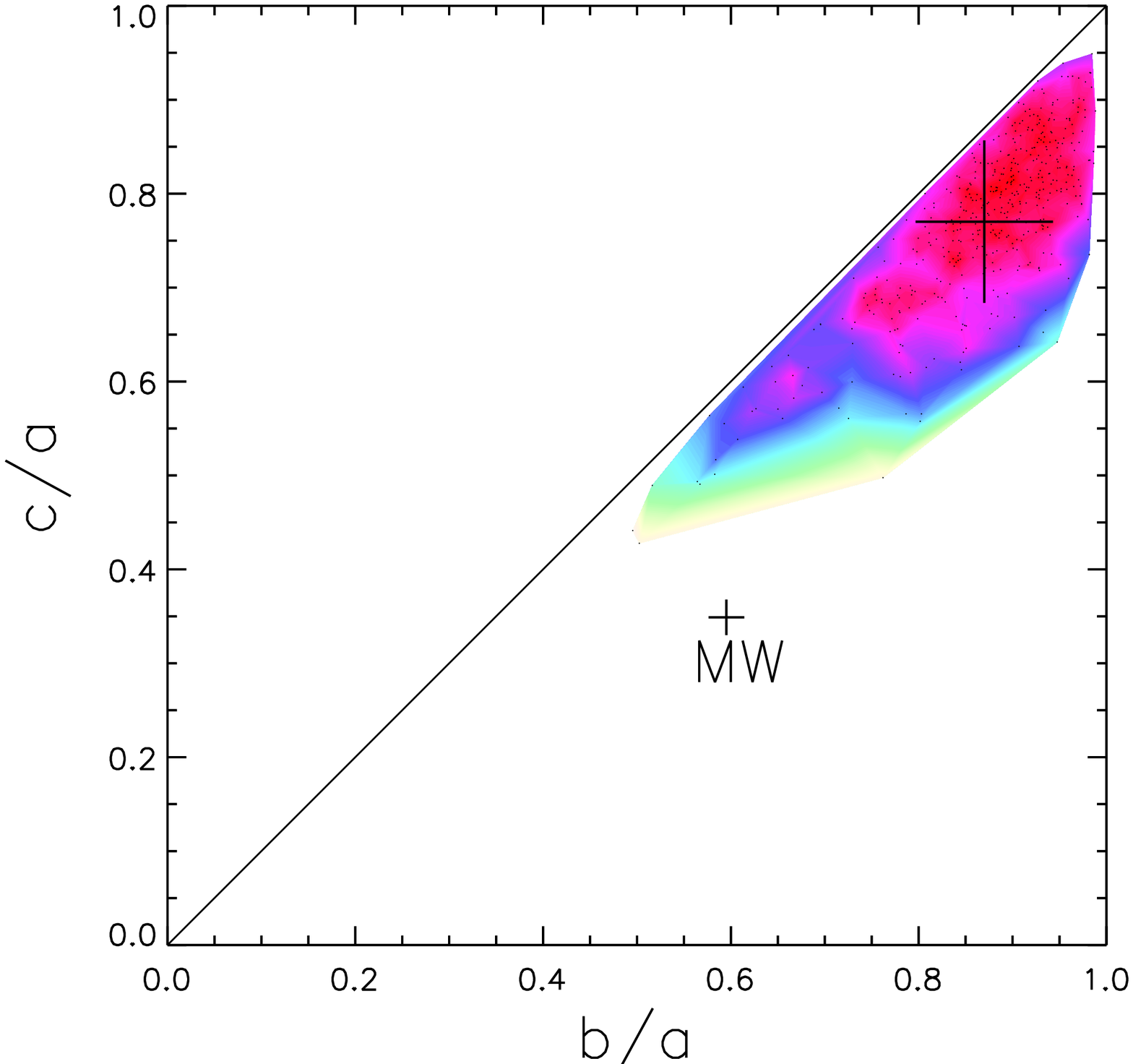}
\caption{The ratio of short-to-long axis {\em versus} the ratio of
intermediate-to-long axis for our halo sample. Each halo is shown as a
black point while contours represent the density of points on an
arbitrary scale, with red being the highest density
and white
the lowest. The large plus sign marks the median value of ($b/a,c/a$)
and its arms represent the 1$\sigma$ extent of haloes in either
direction. The point marked ``MW" indicates the location on such a
plot of the Milky Way's \textit{disc-of-satellites}}
\label{halo.abc}

%\end{figure}

%\begin{figure}
\includegraphics[width=20pc]{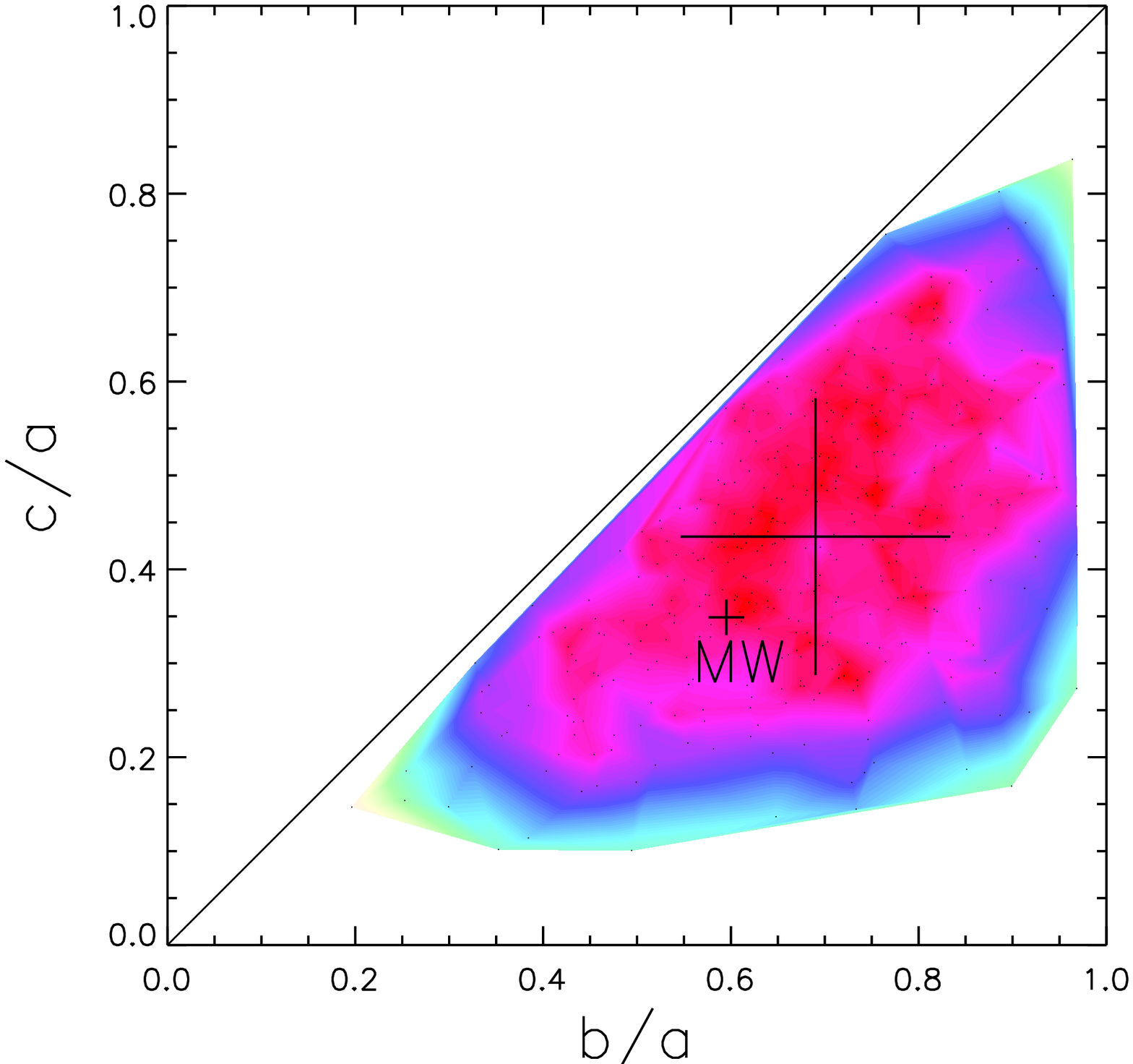}
\caption{Same as Fig.~\ref{halo.abc}, but for our sample of halo satellites.}
\label{sat.abc}
\end{figure}

The flattening of the halos in our sample is evident in
Fig.~\ref{halo.abc} where we plot $c/a$ versus $b/a$. From the
definition of $a$, $b$ and $c$, no points may lie in the triangular
upper half of this plot. The mild asphericity of our sample is
reflected in the tendency for halos to occupy the upper right area of
Fig.~\ref{halo.abc}. The median values of these halo axial ratios are
$<b/a>=0.87$ and $<c/a>=0.77$.

%We find no dependence of halo sphericity on central galaxy colour.

\begin{figure*}

\includegraphics[width=40pc]{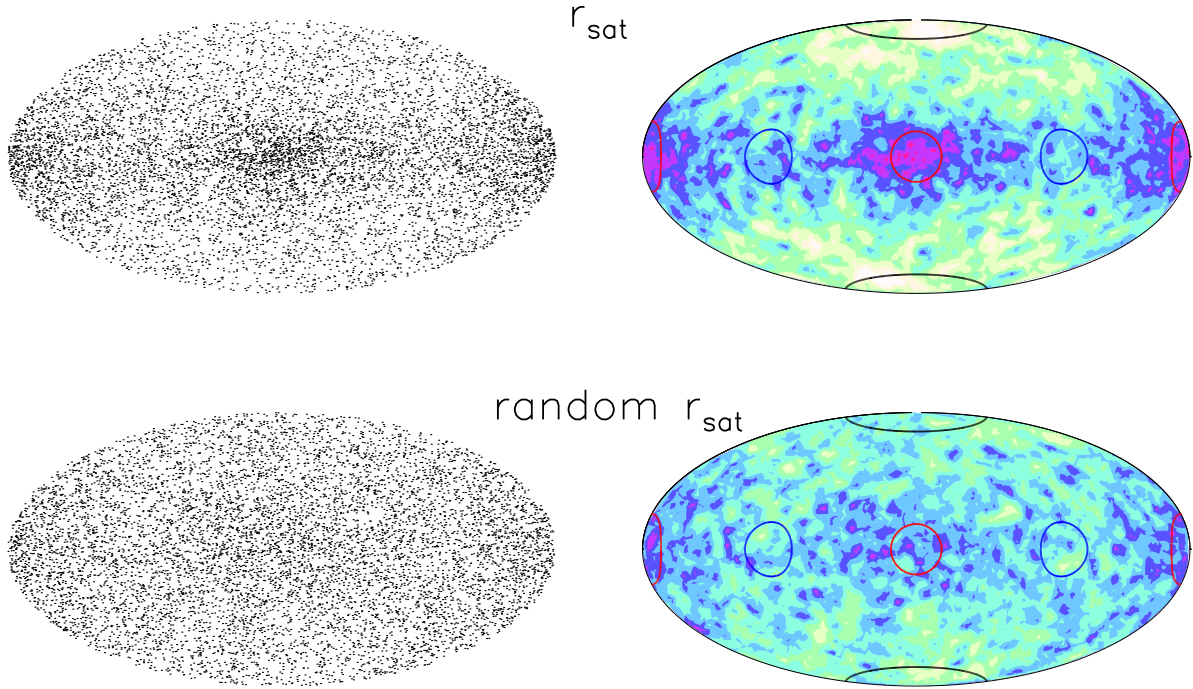}
\caption{\textit{Upper panel}: the location on the sky of all
  satellites in our halo sample, rotated into the frame defined by the
  satellite system's principal axes. On the left we plot the position 
  of each satellite, while on the right we show a false-colour map of
  satellite galaxy ``density'' (defined using the angular distance to
  the 10th nearest neighbor). High density regions are indicated by
  the purple/red shading, while less dense regions are indicated by
  the white shading. The red, blue, and black circles define areas
  within 15 degrees of the $a$, $b$ and, $c$ axes. \textit{Bottom
  panel}: as above but for ``randomized'' satellites.}
\label{sat.pos}
\end{figure*}

The mild asphericity of dark matter haloes may be contrasted with the
much more flattened configurations defined by their satellite systems,
as shown in Fig.~\ref{sat.abc}. This sample shows a much broader
distribution of axial ratios, and contains many more examples of very
triaxial systems than the halo population. The median values of the axial
ratios are significantly smaller than for the halo sample, with
$<b/a>=0.69$ and $<c/a>=0.43$. 

%Again, we find no correlation between central galaxy colour and

%satellite system shape.

The nature of the satellite's aspherical distribution is related to
the fact that satellites do not simply trace the background dark
matter, but occupy biased positions in the halo closer to the
principal axes. We assess the significance of this effect by
conducting a ``randomization'' test of the null hypothesis that
satellites are randomly drawn from the dark matter distribution. For
each satellite in a halo, we randomly select a dark matter particle at
the same radius. We then proceed to calculate the principal axes for
both the satellite systems and those consisting of the random particles
replacing each of the 11 brightest satellites.

We use these two sets of principal axes to project the location of
each satellite and its random counterpart onto a Hammer-Aitoff
projection of the whole sky, shown in Fig.~\ref{sat.pos}. The upper
panels show the location on the sky of the model satellites. For each
central galaxy, we plot the positions of its satellites rotated onto
the reference frame defined by the
\textit{disc-of-satellites} of the 11 brightest satellites. By
definition, the satellites will concentrate along the long axis, a
feature readily visible in the upper panels of Fig.~\ref{sat.pos}.
However, the total fraction of satellites that lie close to the long
axis is determined by the strength of the intrinsic flattening.

We can compare the upper panels of Fig.~\ref{sat.pos} to the lower
panels which show the location on the sky of the random particles
that replaced the brightest satellites projected onto the reference
frame defined by those replacing the 11 brightest in each halo. These
particles also tend to be aligned with the long axis of their
distribution (by construction), but to a lesser extent than the model
satellites. The pseudo-\textit{disc-of-satellites} defined by the
randomly selected dark matter particles is visibly less aspherical
than the true
\textit{disc-of-satellites} defined by the 11 brightest satellites in
each halo.

\begin{figure}
\includegraphics[width=20pc]{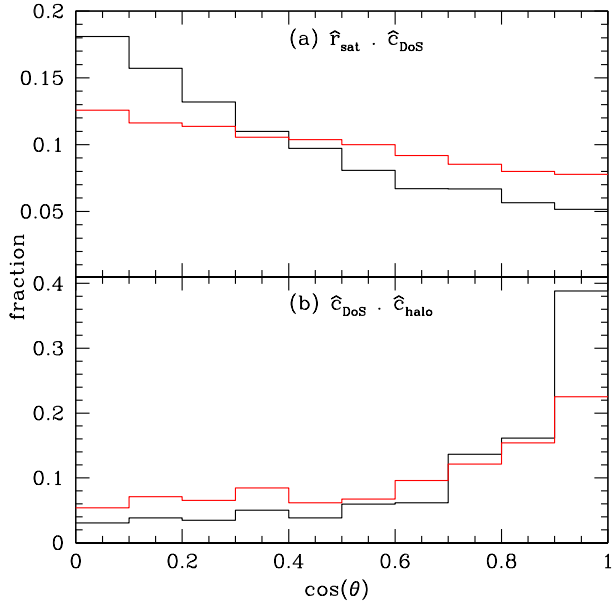}
\caption{\textit{Top panel:} (a) The distribution of the cosine of the angle between the short
axis of the each halo's satellite system ellipsoid and the position of
each satellite galaxy (black curve). The red line shows the analogous
distribution for the ``randomized'' sets of satellite galaxies. \textit{Bottom panel:} (b) The distribution of the angle between the short axis of each halo's dark matter distribution and the short axis of the \textit{disc-of-satellites} composed of the brightest 11 satellite galaxies.}
\label{MW.rc}
\end{figure}

We can quantify this behavior by plotting a histogram of the angle
between the position vector of each satellite ($\hat{r}_{\rm sat}$)
and the short axis (for example) of the ellipsoid defined by the
brightest 11 satellites ($\hat{c}_{\rm sat}$). Fig.~\ref{MW.rc}(a)
shows that the model satellites (black lines) exhibit a much stronger
perpendicular alignment with the short axis of their distribution than
the corresponding set of random dark matter particles (red lines). A
Kolmogorov-Smirnoff test rejects the hypothesis that these two
distributions are drawn from the same parent distribution or from a
uniform distribution at a very high confidence level (formal KS
probabilities $\sim10^{-37}$). Around 15\% more model satellites lie
further than 65 degrees from $\csat$ than expected from the associated
dark matter particles. We conclude that satellite positions are not a
random sample of the halo and that satellite systems are significantly
flatter than their parent halos.

It is straightforward to check the degree of alignment of the
\textit{disc-of-satellites} defined by the 11 brightest galaxies with
their host halo. Fig.~\ref{MW.rc}(b) shows the cosine of the angle
between the short axis of the \textit{disc-of-satellites}
($\hat{c}_{\rm DoS}$) and the short axis of the host halo
($\hat{c}_{\rm halo}$). The model discs-of-satellites (black line)
show a clear tendency to be aligned with the main plane of the
halo. Indeed, nearly 40\% of all satellites systems fall in the last
bin of the histogram. However, a non-negligible fraction of systems
are not aligned, in the sense that their minor axes are nearly
perpendicular to the halo's minor axis. For example, approximately
10\% of the model \textit{discs-of-satellites} are mis-aligned by more
than 75$^\circ$.  To assess the significance of this trend, we define
randomized \textit{discs-of-satellites} using the set of dark matter
particles replacing the brightest 11 model satellites. Again, a
Kolmogorov-Smirnoff test rejects the hypothesis that the actual and
randomized samples are drawn from the same parent distribution, or
that either is drawn from a uniform distribution, at a very high
confidence level.

%Approximately 20\% more modeled systems are within 25 degrees of being

%parallel to their host halos then expected from the matched dark

%matter particle systems. We conclude that in the majority of systems,

%the alignment of the

%\textit{disc-of-satellites} defined by the 11 brightest satellites,

%trace the background dark matter halo's shape.} 

\subsection{Orbital angular momentum of satellites}
A number of recent studies \citep[e.g.][and other references in
\citealt{Metz08}]{Piatek03,Piatek05,Piatek06,Piatek07} have 
obtained upper limits on the proper motion of eight of the Milky Way
(MW) dwarfs. A detailed study of their angular momenta reveal two
particularly interesting aspects of their orbital motion.  The first
is that the angular momentum vectors of at least 6 of these satellites
point to within 30 degrees of the mean angular momentum vector of the
set for which proper motions are available. The second
is that the orbits of at least 5 satellites fall within 30 degrees of
the normal of the
\textit{disc-of-satellites} defined by the 11 brightest dwarfs.

\cite{Metz08} tested for these two properties in  6 
simulations of MW-sized galaxy halos and found that in none of them
did the satellites (identified in a similar manner as in this work)
exhibit the same kind of coherent motion inferred for a subset of MW
satellites. None of the 6 simulated haloes in that work had 5
satellites with with angular momentum vectors within 30 degrees of the
normal to the fitted
\textit{disc-of-satellites}. Just 2 of the halos had 3 satellites
whose angular momentum vectors fell within this region. In addition, 
the mean angular direction of the orbital poles of the simulated
satellites did not lie close to the normal of the 
\textit{disc-of-satellites}, in contrast to the MW.

Metz et al. concluded from this study that the dynamics of at least 5
``rotationally supported'' MW satellites were (at least) hard to
reproduce in CDM simulations. They suggested that the mismatch between
the simulations and the observations could, in theory, be explained if
the satellites of the MW had a different origin from those in the
simulations, for example, if they were ``tidal dwarfs'', formed in a
tidal stream generated in a recent major merger
\citep[e.g.][]{MetzKroupa}.

\begin{figure*}
\includegraphics[width=40pc]{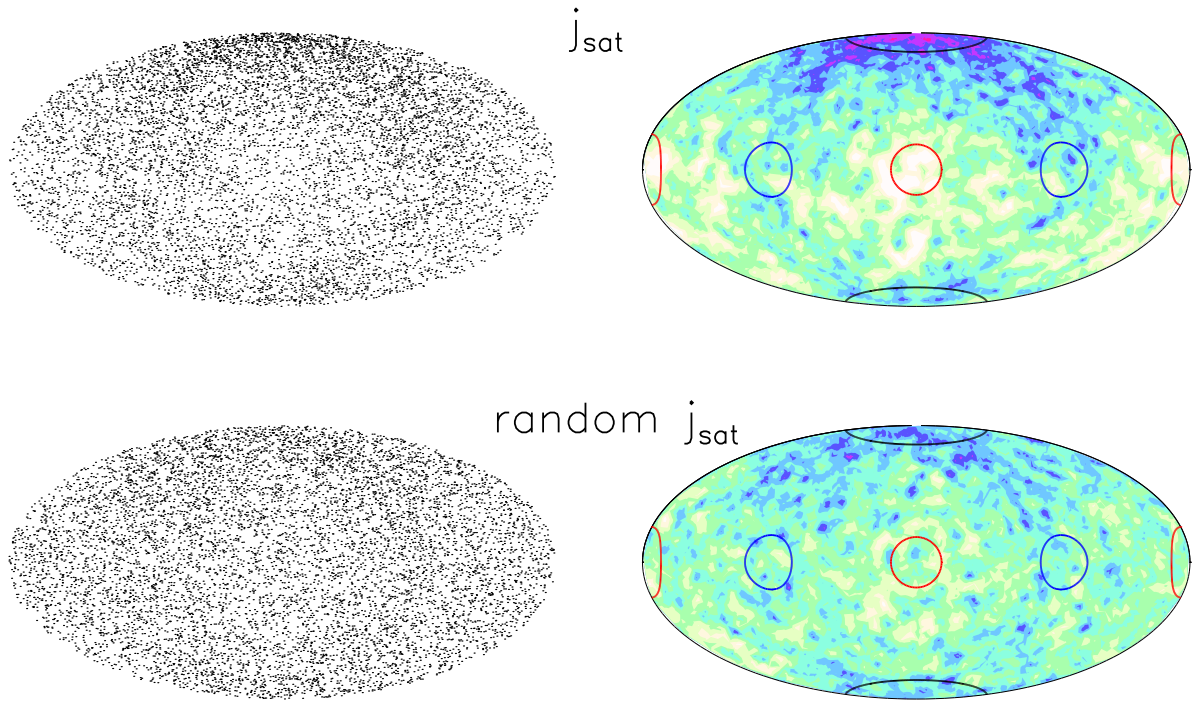}
\caption{\textit{Upper panel}: The location on the sky of the orbital
  angular momentum vector for all satellites in our halo sample,
  rotated into the principal axes of the disc-of-satellites. In these
  projections we choose from the two possible opposing directions for
  the $c$-axis the one which is closest to the mean
  angular momentum vector of the satellite sample. On the left we
  show the location of each satellite's angular momentum vector, while
  on the right we show a contour map of satellite angular momentum
  ``density'' (defined by the angular distance to the 4th nearest
  neighbor). High density regions are indicated by the red/purple
  shading, while less dense regions are indicated by the white
  shading. The red, blue, and black circles define areas within 15
  degrees of the $a$, $b$ and, $c$ axis. \textit{Bottom panel}: The
  same as above, but for randomized satellites.}
\label{sat.ang}
\end{figure*}

Our sample of MW type satellite systems is well suited to testing
whether or not the orbital dynamics of satellites in $\Lambda$CDM
simulations are consistent with the MW data. In Fig.~\ref{sat.ang} we
show the location on the sky of the orbital angular momentum of each
satellite (upper panels) and its randomized counterpart (lower panels)
in our halo sample. We rotate each point on the sky into the frame
defined by the principal axes of the ``\textit{disc-of-satellites}''
defined either by the model satellites or by their randomized
counterparts. There is an ambiguity in choosing which direction of the
$\csat$ axis to take to define the short axis. We choose the $\csat$
axis to point in the direction which is closest to the mean angular
momentum vector of each system. Therefore, the satellites that fall in
the Southern hemispheres of Fig.~\ref{sat.ang} are counter-rotating
with respect to the mean angular momentum of the
\textit{disc-of-satellites}.

Two features are readily visible in this figure. Firstly, the
directions of the angular momentum of the randomized dark matter
particles replacing the model satellites are much more uniform
than those of the actual model satellites. Secondly (by construction),
in both samples the distributions show an excess around the ``North''
pole, which is much more pronounced for the model satellites than for
their randomized counterparts.  On closer examination we note that, as
well as the higher density around the North pole, there is a tendency
for the angular momenta to point close to the $\bsat-\csat$
plane. This effect may be produced by the radially biased orbits of the
satellites in our sample. If a satellite lies along or near $\asat$,
and if it is also moving close to this direction, its angular momentum
vector will point somewhere on the $\bsat - \csat$ plane.

\begin{figure}
\includegraphics[width=20pc]{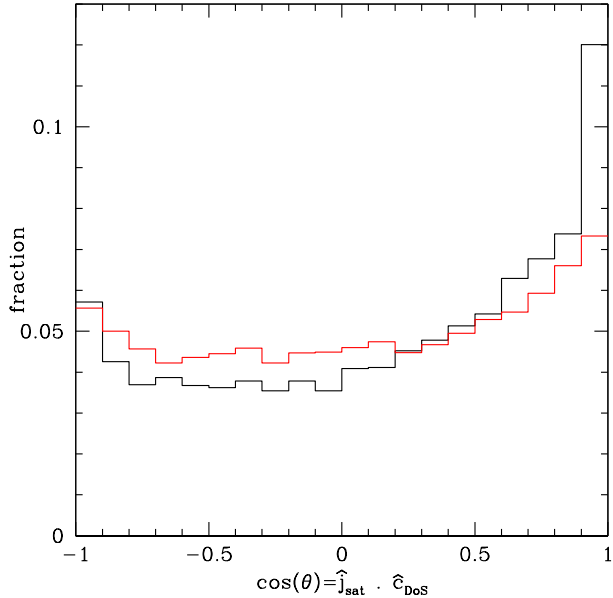}
\caption{The distribution of the cosine of the angle between $\hat{\rm
j}_{\rm sat}$, the satellite galaxy orbital angular momentum, and
$\csat$ (black lines). The red lines show results for the control
sample of randomized dark matter particles.}
\label{MW.jc}
\end{figure}

Satellites have a much stronger tendency to display coherent orbital
angular momenta then the background dark matter. In Fig.~\ref{MW.jc}
we show the distribution of the cosine of the angle between each
satellite's orbital angular momentum vector, $j_{\rm sat}$, and
$\csat$. This plot quantifies what is visually apparent in
Fig.~\ref{sat.ang}: an above random alignment between the orbital
angular momentum of individual satellites and the short axis of the
ellipsoid to which it belongs. The Kolmogorov-Smirnoff
probability that the model satellites and the random particles are drawn
from the same population is negligible.

Thus, our model satellites display the same kind of apparent coherent
motion and spatial flattening as the real MW satellite system. Yet,
the nature of this coherent motion is unclear. Metz et al. suggested
that in the MW, the \textit{disc-of-satellites} is supported purely by
rotation. In this scenario, the majority of satellites move on
circular orbits such that the \textit{disc-of-satellites} itself is
supported by the rotation of the galaxies orbiting in it. In order to
test this hypothesis we carry out the following calculation. As in
Fig.~\ref{sat.ang}, we define the $z$-axis to be parallel to the short
($c$) axis of the satellite distribution and of the two opposing
directions this defines, we select the one closest to the direction of
the mean angular momentum of the satellites.  For each galaxy we calculate
its angular momentum in this `\textit{z}'- direction and call this
quantity $J_{\rm z}(E)$. We then estimate the angular momentum of a
satellite with the same orbital energy, but on a purely circular
orbit, $J_{\rm z c}(E)$. If these two quantities are similar (i.e. $
J_{\rm z }(E) / J_{\rm z c}(E)
\approx 1$), then the satellite's orbit must be approximately
circular and rotationally supported.  Note that since a circular orbit
maximizes the angular momentum, the ratio $ J_{\rm z }(E) / J_{\rm z
c}(E)$ must always be $\leq 1$.

\begin{figure}
\includegraphics[width=20pc]{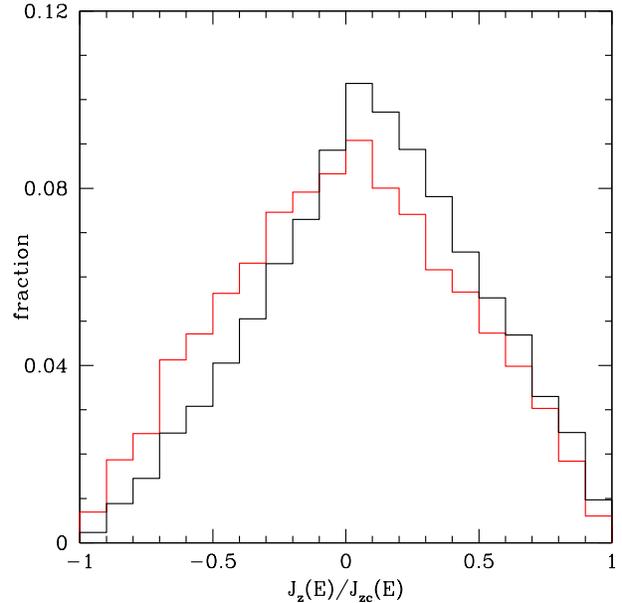}
\caption{Histogram of the ratio of the $z$- component of the angular momentum of each
  model satellite, $J_{\rm z}(\rm E)$, to the angular momentum of a
  satellite on a circular orbit with the same energy, $J_{\rm zc} (\rm
  E)$ (black lines). If the \textit{disc-of-satellites} were rotationally
  supported, all orbits would be circular and the distribution should
be strongly peaked around $\sim$~1. The red lines show the the
histogram for the control sample of matched dark
matter particles.}
\label{jratio}
\end{figure}

For each halo, we estimate its concentration assuming a Navarro, Frenk
\& White (1996; NFW) profile. Once we know the halo concentration, 
we use eqns~(2.18) and~(2.19) of \cite{LaceyCole}, together with the
satellite's radial position in the halo, to calculate its energy. We
then search for the parameters (i.e. the distance and velocity) of a
circular orbit within an NFW potential with the same energy. In this
way, we build the ratio $ J_{\rm z }(E) / J_{\rm z c}(E) $ whose
distribution we show in Fig.~\ref{jratio}.

There are two salient points visible in Fig.~\ref{jratio}. The first
is that the distributions are not peaked at $\approx 1$, but rather at
$\approx 0$, implying that $ J_{\rm z }(E) \ll J_{\rm z c} (E) $. The
majority of the satellites and their randomized counterparts are
therefore {\em not} on circular orbits around the $z$-axis. It is
interesting to note that the control sample exhibits a more pronounced
symmetry about 0 than the satellite sample, implying similar numbers
of counter- and co-rotating particles. The satellite sample, on the
other hand, is somewhat skewed towards positive values of $ J_{\rm z
}(E) / J_{\rm z c}(E)$, implying that more satellites are co-rotating.

We now quantify how common the configuration of the satellite system
of the MW is in our $\Lambda$CDM simulation.  Of the 11 classical
satellites in the MW, 8 (Sagittarius, Draco, Fornax, Ursa Minor,
Carina, Sculptor the SMC and the LMC) have proper motion estimates
which allow us to estimate the direction of their orbital angular
momentum (e.g. see Table~2). 
Of these eight, the LMC, the SMC, and Fornax have angular
momentum pointing well within 30 degrees of the direction to the
normal of the plane fitted to the 11 classical satellites. The angular
momentum vectors of an additional 2, Ursa Minor and Carina, have error
bars large enough that they are compatible with a $< 30$ degrees
misalignment with the pole.  Additionally, there are 3 satellites
without proper motion estimates, implying a minimum of 3 out of 11
($\sim$27\%) and a maximum of 8 out of 11 ($\sim$73\%) satellites with
angular momenta within 30 degrees of the perpendicular to the plane
of the \textit{disc-of-satellites}.

\begin{figure}
\includegraphics[width=20pc]{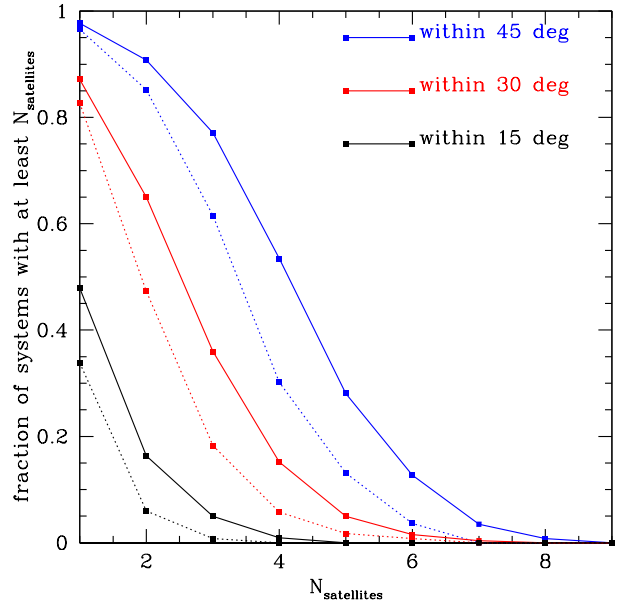}
\caption{The fraction of systems in which the angular momentum of at least
$N$ satellites drawn from the brightest 11 lies within 15 (black), 30
(red), and 45 (blue) degrees of $\csat$ (solid lines) and of $
J_{\rm mean}^{6}$ (dashed lines).}
\label{jincsat}
\end{figure}

To assess how common this form of  alignment  is, we perform the
following calculation: for each simulated halo with more than 11
satellites, we fit the standard plane to the 11 brightest, as
described in Sec.~\ref{halo.oreintation}. We then calculate how often
the orbital angular momentum of \textit{N} randomly selected
satellites falls within 15, 30 or 45 degrees of $\csat$, where the direction of
the $\csat$ axis is chosen, as before, to be the one that is closest
to the mean angular momentum vector. The results are shown in
Fig.~\ref{jincsat}. 

We find, for example, that the probability of finding at least 3
satellites within 30 degrees of $\csat$ is around 35\%, a similar
number to that found in the simulations of Metz et al. If $n=5$, this
value drops to $\sim 5$\%. The probability varies rapidly with
satellite number and the chance of finding 9 satellites within the
specified angle (the total number of classical satellites - save
Sagittarius and Sculptor - that could possibly be aligned) drops to
1\%. The number of satellites with angular momentum aligned with the
normal to the \textit{disc-of-satellites} also depends strongly upon
cone angle. For example, we find 3 satellites within 45 degrees around
70\% of the time but only 5\% of the time within 15 degrees.

\begin{table}
\begin{center}
\caption{Census of the positions and angular momenta of the MW's
satellite population. Column~1
shows the satellite's name. Column~2 gives the galacto-centric
distance of each satellite. Column~3 gives the angular separation
between the normal of the plane fitted to the Milky Way's
\textit{disc-of-satellites} and the angular momentum vector of each satellite.
Column~4 gives angular separation between the direction of 
the mean angular momentum of the 6 satellites
moving in the most ordered fashion, $J_{\rm mean}^{6}$, to 
the angular momentum vector of each satellite.
Columns~5 and~6 indicates whether the uncertainty in the
measurements of the satellite galaxy's angular momentum are consistent
with placing the pole of the angular momentum vector within 30~degrees
of the normal of the \textit{disc-of-satellite's} or within 35~degrees
of $J_{\rm mean}^{6}$. For observational references see
\protect\cite{Metz08} and references therein.}
\begin{tabular}{| p{17mm} | p{5mm} | p{6mm} | p{6.5mm} | p{8.5mm} |  p{8.5mm} | }
%\begin{tabular}{c c c c c}
Name & $d_{\rm GC}$ (kpc) &  $\Delta\phi_{\rm csat}$ (deg) &   $\Delta\phi_{\rm J}$ (deg) & within $\hat{n}_{\rm DoS}$ & within $J_{\rm mean}$\\
   \hline
   \hline
	LMC & 50 & 13 & 7 & YES & YES\\
	SMC & 57 & 21 & 15 & YES & YES \\
	Fornax & 140 & 23 & 25& YES &YES \\
	Ursa~Minor & 68 & 38 & 18 & YES & YES\\
	Carina & 103 & 49 & 61 & YES & YES\\
	Draco & 82 & 69 & 55 & NO & YES\\
	Sagitarius & 16 & 118 & 98& NO &NO\\
	Sculptor & 79 & 136 & 156 & NO&NO\\
	Sextans & 86 & --& --& unknown & unknown\\
	Leo I & 205 & --& --& unknown & unknown\\
	Leo II & 250 & --&--& unknown & unknown\\
    \hline
    \hline
    \end{tabular}
 \end{center}
\label{sattable}
\end{table}

\cite{Metz08} also remarked on the coherence of the angular
momenta of the Milky Way's satellites. They chose to omit 2
satellites (Sculptor and Sagitarius) as their angular momentum vectors
point far from the mean and are clearly not members of a ``common
stream". They then computed the mean angular momentum vector and its
error for the remaining 6 satellites that have measured proper
motions.  Relative to this mean angular momentum direction, they found
that a minimum of 4 (the LMC, SMC, Fornax and Ursa Minor) out of 11
($\sim 36$\%), and a maximum of 9 out of 11 ($\sim82$\%), satellites
fall within the 35~degree error circle.

We perform a similar calculation to determine the probability of such
an alignment.  In order to mimic their procedure, for each halo we
first calculate the net angular momentum of a randomly selected 8
satellites from the brightest 11. Then, after omitting the 2
satellites whose angular momenta are furthest from this mean, we
compute the mean angular momentum vector of the remaining 6.  We refer
to this quantity as $J_{\rm mean}^{6}$. Although
\cite{Metz08} did not use this exact prescription, it is equivalent to
theirs and it results in the same selection for the Milky Way.  We
then calculate how often \textit{N} randomly selected satellites fall
within 15, 30 and 45 degrees of the direction of $J_{\rm
mean}^{6}$. Our results are shown as the dotted lines in
Fig.~\ref{jincsat}

We find that the arrangement in the Milky Way, where 4 satellites have
angular momentum lying within 30~degrees of $J_{\rm mean}^{6}$, occurs
reasonably frequently ($\sim 10$\% of the time). However, we find that
the probability of finding the maximum possible number of satellites
(9) with angular momentum vectors pointing within the even larger
angle of 45~degrees from $J_{\rm mean}^{6}$ is negligible. If future
proper motion measurements for the remaining 3 satellites (Sextans,
Leo I and Leo II) reveal that they too are closely aligned with
$J_{\rm mean}^{6}$, then our results suggest that such a setup would
be extremely unlikely in our $\Lambda$CDM galaxy formation model.

\begin{figure}
\includegraphics[width=20pc]{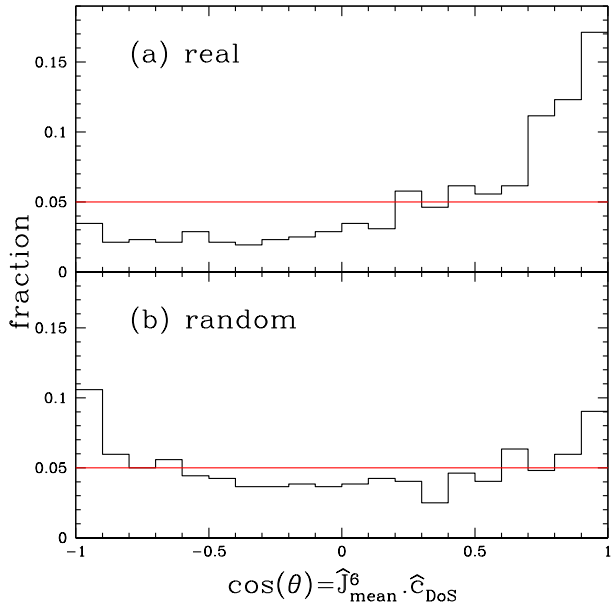}
\caption{The distribution of the angle between the mean orbital
angular momentum, $J_{\rm mean}^{6}$, of the 6 best aligned satellites
and the short axis of the \textit{disc-of-satellites}, $\csat$. We
show the distribution for both the model satellites (a) in the upper
panel, and for the associated randomized dark matter particles (b) in
the lower panel. The dashed red line indicates a uniform distribution
in cos($\theta$).}
\label{jmeanincsat}
\end{figure}

Fig.~\ref{jincsat} shows the probability that the orbital angular
momentum of a random \textit{N} out of 11 satellites falls close to
the short axis of the \textit{disc-of-satellites} or $J_{\rm
mean}^{6}$. The {\em mean} orbital angular momentum of \textit{N}
satellites could still lie close to the axis (as in the MW), even if
the angular momentum of the individual satellites are widely
distributed. In Fig.~\ref{jmeanincsat}a, we show the distribution of
the angle between the mean orbital angular momentum of the 6 best
aligned satellites, $J_{\rm mean}^{6}$, and the short axis of the
\textit{disc-of-satellites}, $\csat$.  

Fig.~\ref{jmeanincsat}a shows that there is an overabundance of
systems that fall in the highest bin, corresponding to an angle of
$\theta \lsim 25$ degrees. The fraction of systems whose $J_{\rm
mean}^{6}$ lies within 25 degrees of $\csat$ is around 17\%, well
above the 5\% expected for a uniform distribution. This is not an
exceedingly large proportion, and although \cite{Metz08} failed to
find a single example in 6 simulations, our result - that such an
alignment occurs roughly 17\% of the time - is consistent with their
findings. There is roughly a 50\% chance of seeing such an alignment
in just 1 out of 6 systems drawn at random from our sample. Yet, the
non-uniformity of the distribution of the angle between $J_{\rm
mean}^{6}$ and $\csat$ is statistically very significant: a
Kolmogorov-Smirnoff test rules out the hypothesis that this sample is
drawn from a random distribution at very high confidence.

We contrast these results with the histogram in the lower panel of
Fig.~\ref{jmeanincsat}b which shows the distribution of the same angle
but now with  $J_{\rm mean}^{6}$ defined by the matched randomly
selected control particles. In this case, the distribution is
consistent with uniformity: the KS probability of drawing such a
sample from a uniform distribution is roughly 20\%. In other words,
the orbital angular momentum of the control particles (which were
chosen to be at the same radial distances as the model galaxies),
exhibits no preference for pointing near the normal of the plane they
define.

\section{Conclusion and Discussion}
In this paper we have used a large cosmological $N$-body simulation of
the $\Lambda$CDM cosmology, populated with galaxies using the \textsc{
Galform} semi-analytic model of galaxy formation, to investigate the
expected frequency of the peculiar alignments exhibited by the
satellite galaxies of the Milky Way. We confirm previous work
\cite[e.g][]{Kang05, Libeskind05, Zentner05} which found highly
flattened spatial distributions of satellites within the much more
spherical dark matter distributions of simulated galactic halos. The
satellite ellipsoid tends to be aligned with the shape of the dark
matter halo, although in a small fraction of cases it is
anti-aligned.

%We found no central galaxy colour dependency of the flattening.

We have also used our simulation in an attempt to understand the
dynamics of satellites within Milky Way type halos. We have found that
systems that resemble the Milky Way - with at least 3 satellites
orbiting within 30 degrees of the normal to the plane defined by the
11 brightest of them - occur roughly 35\% of the time. This is
consistent with the work of \cite{Metz08} who found 2 out of 6
galaxies showing this behavior. However, the fraction of systems
which have 8 satellites all orbiting with angular momentum aligned to
within 30 degrees of the normal to their plane is less then 1.5\% at
the 95\% confidence level. Thus, if reliable proper motion estimates
for the remaining 5 ``classical" Milky Way satellites reveal that they
too orbit near this plane, this coincidence would be difficult to
explain within our $\Lambda$CDM galaxy formation model. In roughly
17\% of our simulated galactic halos we find systems in which the {\em
mean} orbital angular momentum of the 6 best aligned satellites lies
within 25~degrees of the short axis of the satellites' plane (and, in
30\% of cases within 35~degrees), as seen in the Milky Way.

The aligned satellites in the simulations do not make up a
rotationally supported \textit{disc-of-satellites}. Similarly, and
contrary to the suggestion by \cite{Metz08}, the rough alignment of
some of the MW's satellites with the pole of their
\textit{disc-of-satellites} does not imply that the system is
rotationally supported. This can only be ascertained by comparing the
angular momenta of the satellites with the values for circular orbits
of the same energy. Only this comparison (which requires various model
assumptions) can expose how circular the orbits really are. If indeed
the orbits of the Milky Way satellites turn out to be rotationally
supported, this would be difficult to explain in the $\Lambda$CDM model.

The main result of this paper is the robust quantification of the
fraction of Milky Way type galaxies in the $\Lambda$CDM cosmology that
resemble our Galaxy as far as its anisotropic satellite distribution
and dynamics are concerned. While previous analysis of simulations 
\citep{Libeskind05,Libeskind07,Zentner05,Metz08} have come to 
some of the same conclusions as we do here, their statistical
significance was compromised by small sample sizes (6, 3, 3 and~6 haloes
respectively). Our large sample of over 400 Milky Way sized haloes
allows us robustly to quantify the expected frequency of both
satellite flattening and coherent satellite motion.

Future surveys such as Pan-STARRS or  SkyMapper
\citep[]{Kaiser02,Keller07} 
will obtain a more complete census of the Milky Way's satellite
population, including measurements of the galaxy luminosity function,
and better determinations of their  kinematic properties and their 
relationship with the ``great pancake'' of bright satellites. 
It was originally suggested by
\cite{Benson02}  \citep[see also][]{Koposov07, Tollerud08}, that new surveys may reveal hundreds, if not
thousands, of faint small satellites. However, it seems highly
unlikely that the known population of bright Milky Way satellites 
will change significantly and thus that the conclusions of
this paper -- which pertain to the bright satellites -- will be
significantly altered.

\section*{Acknowledgments}
NIL is supported by the Minerva Stiftung of the Max Planck Gesellschaft.

\end{document}